\documentclass[english]{emulateapj}
\usepackage[T1]{fontenc}
\usepackage[latin1]{inputenc}
\usepackage{graphicx}
\usepackage{amssymb}

\providecommand{\tabularnewline}{\\}
\newcommand{\lyxdot}{.}

\usepackage{graphicx}
\usepackage{amssymb}
\usepackage{amsmath}
\usepackage{times}

\shorttitle{The asymmetric distribution of HVSs}
\shortauthors{Perets et al. }

\usepackage{babel}

\begin{document}
\global\long\def\Mo{M_{\odot}}
\global\long\def\Ro{R_{\odot}}
\global\long\def\Lo{L_{\odot}}
\global\long\def\SgrA{\mathrm{Sgr\, A^{\star}}}
\global\long\def\Ms{M_{\star}}
\global\long\def\Mbh{M_{\bullet}}
\global\long\def\rMP{r_{\mathrm{MP}}}
\global\long\def\aGW{a_{\mathrm{GW}}}

\title{The Galactic potential and the asymmetric distribution of hypervelocity
stars}

\author{Hagai B. Perets$^{1}$, Xufen Wu$^{2}$, HongSheng Zhao$^{2}$, Benoit
Famaey$^{3}$, Gianfranco Gentile$^{3,4}$ and Tal Alexander$^{1,5}$}

\email{hagai.perets@weizmann.ac.il}

\affil{$^{1}$ Weizmann Institute of Science, POB 26, Rehovot 76100, Israel\\
$^{2}$Scottish University Physics Alliances, University of St
Andrews, KY16 9SS, U.K.\\
$^{3}$Institut d'Astronomie et d'Astrophysique, Universite Libre
de Bruxelles, Boulevard du Triomphe, B-1050 Bruxelles, Belgium\\
$^{4}$Sterrenkundig Observatorium, Ghent University, Krijgslaan
281, S9, B-9000 Ghent, Belgium\\
$^{5}$The William Z. and Eda Bess Novick career development chair}
\begin{abstract}
In recent years several hypervelocity stars (HVSs) have been observed
in the halo of our Galaxy. Such HVSs have possibly been ejected from
the Galactic center and then propagated in the Galactic potential
up to their current position. The recent survey for candidate HVSs
show an asymmetry in the kinematics of candidate HVSs (position and
velocity vectors), where more outgoing stars than ingoing stars (i.e.
positive Galactocentric velocities vs. negative ones) are observed.
We show that such kinematic asymmetry, which is likely due to the
finite lifetime of the stars and Galactic potential structure, could
be used in a novel method to probe and constrain the Galactic potential,
identify the stellar type of the stars in the survey and estimate
the number of HVSs. Kinematics-independent identification of the stellar
types of the stars in such surveys (e.g. spectroscopic identification)
could further improve these results. We find that the observed asymmetry
between ingoing and outgoing stars favors specific Galactic potential
models. It also implies a lower limit of $\sim54\pm8$ main sequence
HVSs in the survey sample ($\gtrsim648\pm96$ in the Galaxy), assuming
that all of the main sequence stars in the survey originate from the
Galactic center. The other stars in the survey are likely to be hot
blue horizontal branch stars born in the halo rather than stars ejected
from the Galactic center. 
\end{abstract}


\section{Introduction}

Hypervelocity stars (HVSs) are stars with extremely high peculiar
velocities relative to the velocity distribution of their parent population.
In recent years several HVSs have been observed in the Galactic halo,
some of them unbound to the Galaxy (with velocities beyond the escape
velocity;\citealp{bro+07b}). From these a Galactic population of
$96\pm10$ such HVSs was inferred \citep{bro+07b}, up to the $100$
kpc distance limit of the survey. Many similar bound HVSs (with velocities
lower than the escape velocity) have been observed at larger numbers.
Most of the observed HVSs are B-type stars (\citealt{bro+05,bro+06a,bro+06b,bro+07a,bro+07b,ede+06};
future observations of HVSs of other stellar types are discussed in
\citealt{kol+07,bro+08,ken+08}). Given the color selection of the
targeted survey for these stars \citep{bro+06a}, such stars could
be either main sequence (MS; or blue straggler) B stars or hot blue
horizontal branch (BHB) stars. Currently only three of the stars in
the survey have specific unambiguous identification and were found
to be MS stars \citep{fue+06,lop+08,prz+08b}. 

Extreme velocities as found for these stars most likely suggest a
dynamical origin from an interaction with or close to the massive
black hole (MBH) in the Galactic center \citep[GC; ][]{hil88,yuq+03,lev05,ole+07,per+07}.
In the following we discuss only HVSs ejected from the GC%
\footnote{\label{fn:Disk HVSs}Note that recent observations possibly suggest
a different origin for two of HVSs observed serendipitously (\citealp{bon+08,prz+08,heb+08},
but see \citealp{per08b}). Throughout this paper we assume that most
(if not all) of the MS stars in the HVSs survey sample have a GC origin.
We note however that the methods we describe here to constrain the
Galactic potential could similarly be used, in principle, for the
analysis of high velocity stars ejected from the Galactic disk. %
} and observed in the Galactic halo, $>10$ kpc from the GC, which
would require ejection velocities from the GC of $\gtrsim800$ km
s$^{-1}$. Such HVSs could serve as probes of the GC environment,
stellar population and dynamics (see e.g. \citealp{ses+07a,ken+08,per08a})
and serve as an independent evidence for the existence of a MBH in
the Galactic center \citep{hil88}. Most of the B-type stars observed
through the HVSs survey have lower velocities and are either bound
HVSs \citep{bro+07a} or are just halo stars. The survey shows an
asymmetry in the kinematics of the stars, where more stars have positive
radial velocities in Galactocentric coordinates (i.e. outgoing stars)
than negative ones (ingoing or returning stars). As we show this asymmetry
is dependent on both the absolute velocity and the distance of the
stars from the GC, and could be used in a novel method to probe and
constrain the Galactic potential, identify the stellar type of the
stars in the survey and estimate the number of HVSs.

This paper is organized as follows. We first briefly describe the
HVSs survey (\S \ref{sec:vel-dist}), and then suggest a novel method
to probe the Galactic potential using such surveys (\S \ref{sec:HVS-potential}).
In \S \ref{sec:HVS-lifetime} we use a similar analysis to infer
the statistics of the stellar type of stars in the HVSs survey and
estimate a lower limit to the number of HVSs in the Galaxy.

\section{The velocity-distance distribution of HVSs}

\label{sec:vel-dist}

HVSs of almost any stellar type could theoretically be observed, since
the currently suggested scenarios for the origin of HVSs give rise
to only limited number of constraints on their stellar characteristics
(e.g. \citealp{han07,per08a}). However, given their relatively small
numbers in the Galaxy, it is practically impossible to find HVSs close
by. For this reason, following the discovery of the first HVSs in
the Galactic halo \citep{bro+05,hir+05,ede+06}, \citet{bro+06b}
have issued a survey of HVSs extending to large distances. They have
searched for HVSs among color selected B-type halo stars of limited
magnitude (a more recent survey also searches for A type stars; \citealt{bro+08};
not included in our analysis). Such stars are luminous enough to be
observed at large distances in the Galactic halo where the relative
frequency of HVSs is much higher, and are less likely to be contaminated
by the disk and halo stellar population. These stars could be either
MS B stars (or blue stragglers) with masses of $3-4\, M_{\odot}$and
short lifetimes ($1-4\times10^{8}$ yrs) or hot blue horizontal branch
(BHB) stars also with short lifetimes (few$\times10^{8}$ yrs on the
horizontal branch), but long progenitor lifetime. The absolute magnitudes
of observed stars and hence their inferred distances depend on their
stellar type. Any analysis of the distribution of HVSs should take
both possibilities into account. 

Fig. 2 in \citet{bro+07b} shows a definite asymmetry in the distribution
of ingoing and outgoing HVSs, where the velocities of ingoing stars
do not extend beyond $300$ km s$^{-1}$. Given that the escape velocity
at these distances is much higher, one would expect to see bound HVSs
returning at velocities up to the escape velocity, in striking contrast
with observations. In order to study this behavior, we turn to the
velocity-distance distribution of these stars. In fig. \ref{f:MS-HVSs}
we show the radial velocity-distance distribution (relative to the
GC) for all of the observed B-type stars in the \citet{bro+07b} survey,
assuming they are either MS stars (and are therefore more luminous
and more distant; fig. \ref{f:MS-HVSs}a), or hot BHB stars (and therefore
closer; fig. \ref{f:MS-HVSs}b).

\section{Probing the Galactic potential using hypervelocity stars }

\label{sec:HVS-potential}

Many studies have been done to constrain the Galactic potential at
large distances through observations \citep[see e.g. ][and references therein]{fic+91,bat+05,smi+07,xue+08}.
Some of these studies use the velocity dispersion of observed objects
to constrain the Galactic potential \citep[e.g. ][]{bat+05,xue+08},
however these suffer from uncertainties regarding the velocity anisotropy
and the behavior of the stellar halo density at very large distances,
and require some apriori assumptions regarding these parameters, which
may lead to large uncertainties \citep[see, for example, discussion in ][]{deh+06}.
In additions many objects are needed in order to obtain the velocity
dispersion at a given distance from the GC. Other studies explore
the local escape velocity from the Galaxy through observations of
high velocity stars \citep{smi+07}. However, such analysis contains
degeneracies and depends on the unknown structure of the tail of the
velocity distribution of the high velocity stars. Consequently specific
assumptions must be taken for the velocity distribution, which could
be strongly affected by the small number statistics of the observed
highest velocity stars in the distribution tail. Moreover, the assumptions
used for the stellar velocities depend on their being extended up
to the escape velocity from the Galaxy. Although large surveys may
help solve this problem, very high velocity stars are quite rare in
the Galaxy, and would be difficult to find especially in surveys limited
to relatively close environment of the solar neighborhood. 

\citet{gne+05} and \citet{yuq+07b} suggested to use the kinematics
of HVSs in order to probe the Galactic potential using the position
and velocity vectors of HVSs at large Galactocentric distances. Under
the assumption that HVSs were ejected from the GC they suggest to
measure the slight departure from purely radial orbits of these HVSs,
due to the (possible) triaxiality of the Galactic potential. These
methods require the accurate distance and 3D velocity of HVSs, and
focus on the triaxiality of the Galactic potential, which is important
in the context of hierarchical, cold dark matter (CDM) models of structure
formation \citep[e.g. ][]{hay+07}, although we note this should be
interesting also in respect to modified Newtonian dynamics (MOND)
theories \citep{mil+83a}. Recently, \citet{ken+08} have studied
the propagation of HVSs in the Galactic potential and showed that
it could depend strongly on the Galactic potential at the central
regions of the Galaxy ($200$pc). They also showed the dependence
of the HVSs radial distributions on the stellar type, and their observational
implications. 

In the following we suggest a method which is more general in nature
and more useful in constraining and distinguishing between different
Galactic mass distributions which are required by the different Galactic
potential models at large distances (although it could also be relevant
to probing the triaxiality of the Galactic potential). This can also
be used in order to discriminate between different CDM Galactic potential
models and/or between Galactic potentials in MOND theories. This method
makes use of the asymmetry in velocity distributions of ingoing and
outgoing HVSs (see also related discussion in \citealp{ken+08}).
We begin with a naive description of the method, assuming one could
observe even the oldest bound HVSs (i.e. those that could not leave
the Galaxy, and could have gone through the Galaxy a few times). We
then continue with a more realistic treatment which takes into account
the finite lifetime of stars observable in the halo, given the limited
observational capabilities.

\subsection{Long lived, observable hypervelocity stars}

Let us assume that HVSs have been continuously ejected from the GC
with some distribution of velocities, which would produce both bound
and unbound HVSs. Unbound stars eventually leave the galaxy. Bound
stars reach tha apo-apse point of their orbit and then return back
to the GC with negative radial velocity (in Galactocentric coordinates).
The Galactocentric distance-velocity distribution of ingoing stars
would then have a cut-off, which would correspond to the escape velocity
of these stars at a given distance from the GC. Such a cut off is
distance dependent and thus more distant HVSs would have lower absolute
velocities. At the same time we should see that the distribution of
outgoing HVSs extends to much higher absolute velocities, since this
population includes the unbound stars, on their way out of the galaxy.
Consequently, a clear asymmetry should be observed between the distribution
of ingoing and outgoing stars. This asymmetry or cut-off in the distance-velocity
distribution would map the escape velocity of stars from the galaxy
at any given distance where HVSs are observed, and serve as a direct
probe of the galactic potential. Note, however, that very different
galactic potential may have escape velocities which are quite similar
at a wide range of distances from the GC \citep{wu+08}. In such cases
observations of more distant HVSs (i.e. wider distance range) may
be required to distinguish between such potentials. Since such stars
would be fainter, this would be more difficult observationally. 

The use of ingoing and outgoing halo HVSs has three advantages over
methods used to probe the local escape velocity from the Galaxy. First,
there is a clear natural separation between bound and unbound stars,
and the latter can not contaminate the sample of ingoing high velocity
bound stars which are used to calculate the escape velocity. Second,
there is no required assumption regarding the structure of the velocity
distribution and its tail. Third, the HVSs are observed over a large
distance range, and could thus map the Galactic potential in this
full range, and not only at the local scale as have been done with
high velocity stars in surveys such as the Rave survey \citep{smi+07}.

\subsection{Realistic short lived hypervelocity stars}

In reality, observable stars in the HVSs targeted survey may not have
an unlimited propagation time, as was our naive assumption. This may
result either because of their short lifetimes (after which they evolve
to a different stellar type, which can not be observed at such distances
with current instruments) or due to their possible origin from a burst
like event, which ejected HVSs only over a limited short time, and
not as a continuous process occurring over the lifetime of the Galaxy
(see \citealt{per08a})%
\footnote{\label{fn:burst 2}In such a case, for example, the highest velocity
bound HVSs may never be observed as ingoing HVSs, since a longer propagation
time is required for them to reach the apo-apse point of their orbit
and become ingoing HVSs.%
}. Nevertheless, the general method prescribed above could still be
applicable, with some modifications. In fact, as we the limited propagation
time of HVSs may prove to be more advantageous in some respects). 

Assuming some finite propagation time for HVSs, we would still expect
an asymmetry in the ingoing and outgoing HVSs distance-velocity distributions.
However, in this case the cutoff in the ingoing HVSs distribution
would be at much lower velocities than the escape velocity. This cutoff
corresponds to the maximal return velocity of HVSs which could still
be observed coming back during their short propagation time. Assuming
a maximal propagation time for the HVSs (its lifetime), this cutoff
could thus be used as a probe of the Galactic potential in the same
way as the naive method outlined above. Moreover, the short propagation
time of stars can be advantageous for our purposes. Stars of different
stellar types have different maximal lifetimes and would produce different
distance-velocity cutoffs. Consequently these different populations
can supply us with several independent probes of the Galactic potential,
that, combined together, would further assist in constraining the
Galactic potential. a higher velocity at a given distance implies
a longer travel time (as the travel time depends on the potential
out to the apo-apse of the orbit), ingoing HVSs provide information
not only on the escape velocity at the point where they are observed
today, but even further away. Furthermore, up to the distance-velocity
cutoff the distribution of outgoing and ingoing stars (with lower
velocities than the cutoff velocity at their position) could be compared$^{2,}$%
\footnote{\label{fn:burst ejection}This, however, is true only under the assumption
of a continuous and constant ejection rate, which is not the case
for burst ejection of HVSs by an inspiralling IMBH \citep{lev05,bau+06,loc+07,ses+07b}%
}, in terms of the number of stars; the two dimensional distance-(absolute)
velocity distribution and the stellar types (if known) or color-color
distributions of the samples of outgoing and ingoing stars. Any difference
between these distributions is due to the further propagation in the
Galactic potential of the ingoing HVSs to the apocenter of their orbit
and back. Different Galactic potential models give different return
times, that are also highly sensitive to the velocity of the HVSs.
Consequently, the statistical correlation between the ingoing and
outgoing distance-velocity distributions could serve as a quantitatively
sensitive method for discriminating between models for the Galactic
potential than just the population of the highest velocity ingoing
HVSs.

\subsection{Discriminating between Galactic potential models}

\label{sub:potentials}

\begin{figure}
\begin{tabular}{c}
\includegraphics[clip,width=0.9\columnwidth]{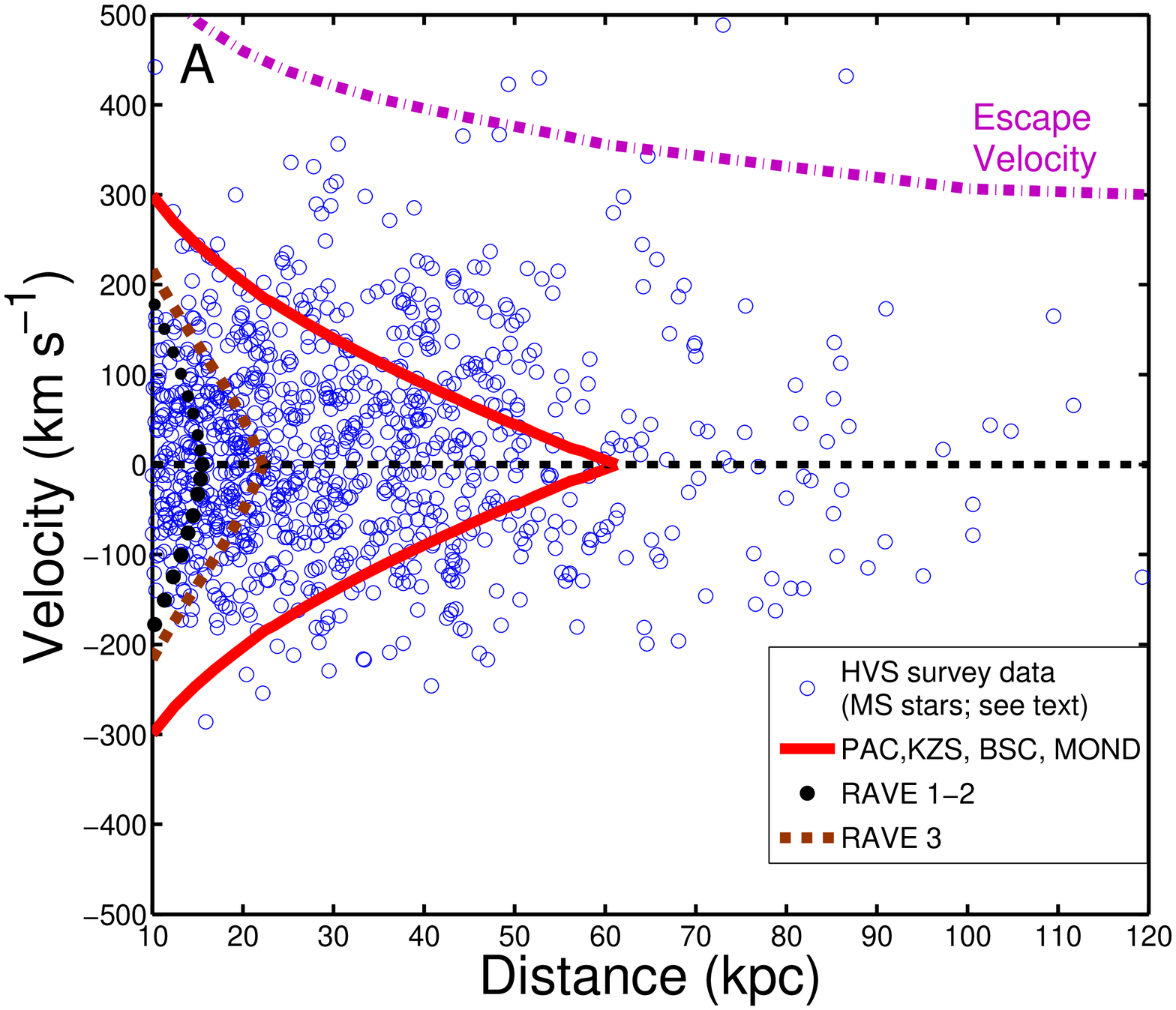}\tabularnewline
\end{tabular}

\begin{tabular}{c}
\includegraphics[clip,width=0.9\columnwidth]{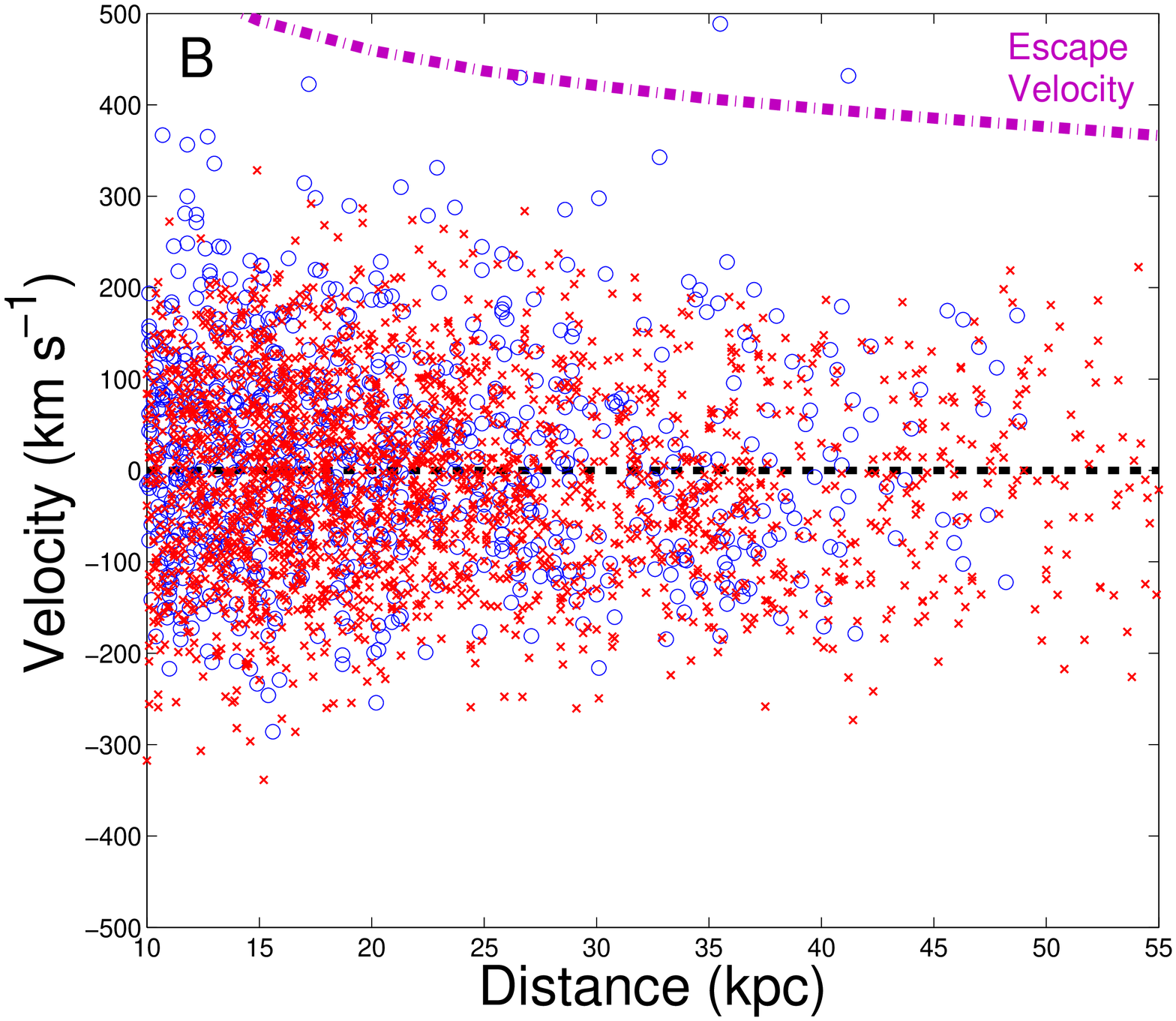}\tabularnewline
\end{tabular}

\caption{\label{f:MS-HVSs}The distance-velocity distribution of Halo B-type
stars from \citet{bro+07b} observations (blue circles). (a) The observed
stars are assumed to be main sequence stars. The lines represent the
critical asymmetry lines (see text) for various potentials (see legend),
for propagation up to the maximal lifetime of such stars ($\sim4\times10^{8}$yrs).
(b) The same, but now assuming all the stars are hot BHB stars, and
therefore less distant. The velocity-distance distribution of regular
halo BHB stars ($\times$marks; taken from \citealp{xue+08}), is
shown for comparison. Dashed middle line in both panels separates
between outgoing and ingoing stars. The escape velocity (in the KZS
model) is also shown for comparison in both panels. }

\end{figure}

Currently, only three HVSs in the HVSs survey have an unambiguous
stellar type identification, and the analysis suggested here for probing
the Galactic potential can not be used directly. The clear cut-off
in the ingoing stars distribution would not be observed, as it would
be smeared by the existence of hot BHB stars contaminating the sample.
Nevertheless we still expect a statistical asymmetry between the number
of ingoing vs. outgoing stars in the sample, which should be observable
beyond the theoretical cut-off. As an illustrative example we show
the critical asymmetry lines for propagation of MS B stars HVSs up
to a maximal lifetime of $4\times10^{8}$. In fig. \ref{f:MS-HVSs}a
we show the critical asymmetry lines for such time limited propagation
in different models for the Galactic potential. The different models
we use include five dark matter (CDM) potentials and one MOND potential.
Beside the \citet{pac+90} model (hereafter PAC) all models are described
in detail in \citet{wu+08}, where the same reference names are used
(KZS, BSC, RAVE1-3 and MOND). The PAC, KZS, BSC and the MOND models
are almost indistinguishable in this range of distances, where as
the RAVE 1-2 (indistinguishable in this range) and RAVE 3 models show
very different behavior (see appendix for a short discussion on the
differences between these Galactic potential models). We look for
asymmetry by counting the numbers of outgoing vs. ingoing stars in
the sample, for the different potentials, i.e. counting the number
of stars above and below the positive and negative velocity lines
of the critical asymmetry curves shown in fig. \ref{f:MS-HVSs}a.
We look for the best fit model, which should show the largest asymmetry.
We find a total of $166$ outgoing stars vs. $112$ ingoing stars
for the PAC,KZS, BSC and MOND models (the probability for getting
such an asymmetric distribution from an apriori symmetric distribution
is $p=10^{-3}$) and $378\,(452)$ outgoing stars vs. $319\,(388)$
ingoing stars for the RAVE 3 (RAVE 1-2) models ($p=0.025\,(0.027)$)
. Although all models show asymmetric distributions, the stars counted
for the RAVE models contain the contribution from the PAC/KZS/BSC/MOND
models. When subtracting this contribution we find that the RAVE models
do not show any additional asymmetry {[}$212\,(286)$ outgoing vs.
$207\,(276)$ ingoing stars; $p=0.8\,(0.67)${]}. Therefore the HVSs
sample favors the PAC,KZS, BSC and MOND models over the RAVE models. 

As we have shown even the contaminated sample of HVSs could already
constrain Galactic potential models. Future identification of the
stellar types of the stars in the survey which could purify it could
give even stronger constraints on these and other Galactic potential
models. In the appendix we give a simple example for the use of such
future data, using mock simulated data of ejected HVSs We also note
that stars with longer MS lifetimes could probe larger distance range
during their propagation and still return during their MS lifetimes
to be observed as ingoing HVS %
\footnote{In fact rejuvenated blue straggler stars could therefore contaminate
the sample of the HVSs, however this population is not expected to
be large \citep{per08b}. %
}. Such stars could therefore serve as observable probes of the Galactic
potential at even larger distances not accessible in any other ways
(and could possibly discriminate between CDM and MOND models that
differ only at these distance ranges). This would require the identification
of later-type HVSs among halo stars (e.g. \citealt{bro+08}, not analyzed
here).

\section{Propagation in the Galactic potential and the lifetimes of observed
hypervelocity stars}

\label{sec:HVS-lifetime}

The method described above uses the finite lifetimes of stars and
their kinematics to constrain the Galactic potential . A very similar
approach could also be used to constrain the number of the HVSs and
their lifetimes, given a specific Galactic potential. Making use of
this approach we show that most of the stars in the HVSs survey, especially
with ingoing velocities are likely to be halo hot BHB stars and have
not been ejected from the GC \citep[see also][for related discussions]{kol+07,yuq+07b,bro+07b,ken+08},
but nevertheless the number of HVSs ejected from the GC could be much
higher than previously thought. 

In the previous section we described the critical distance-velocity
asymmetry lines. For a given Galactic potential and a given propagation
time of a HVS, one could find the critical line outside which no such
ingoing stars should be observed. In other words any ingoing star
beyond this line can not be a HVS from the GC with such (or shorter)
lifetime. We can therefore identify at least some of the HVSs sample
stars as stars that are not MS stars ejected from the GC using this
criteria (see \citealp{sve+07} for a related discussion). Moreover,
since the distribution of such stars should be symmetric (as observed
for other samples of halo objects, such as regular halo BHB stars)
any asymmetry beyond the critical lines is due to the outgoing MS
HVSs from the GC (or from the Galactic disk$^{1}$, a possibility
which we do not discuss here). 

Assuming the PAC model (or the KZS and BSC model that give similar
results) and a maximal propagation time of $4\times10^{8}$yrs (lifetime
of a $3\, M_{\odot}$ MS B star) we find an overabundance of $166-112=54$
outgoing stars, where an asymmetry of $8$ stars correspond to the
1 $\sigma$ probability level. We therefore give a lower limit estimate
for the number of HVSs beyond the critical line of $\gtrsim54\pm8$
HVSs, from which we infer, following the calculations by \citet{bro+07b},
that a total number of $\gtrsim648\pm96$ such stars (main sequence
B stars of $3-4\, M_{\odot}$at distances of $10\, kpc\lesssim r\lesssim100\, kpc$
ejected from the Galactic center) exist in the Galaxy. We expect that
most if not all of the ingoing stars beyond the critical line are
hot BHB stars, where outgoing stars beyond the line could be both
hot BHB stars or MS stars, with a ratio of $2$ to $1$ ($166-54=112$
vs. $54$). These large estimated number of HVSs may suggest a different
contamination from high velocity stars ejected from the Galactic disc,
that could also produce an asymmetric distribution due to finite MS
lifetimes of the stars. Here we do not address this possibility, which
requires a more detailed study and would be addressed in a another
paper (Perets el al., in preparation). 

\begin{figure}
\begin{tabular}{c}
\includegraphics[clip,width=0.9\columnwidth]{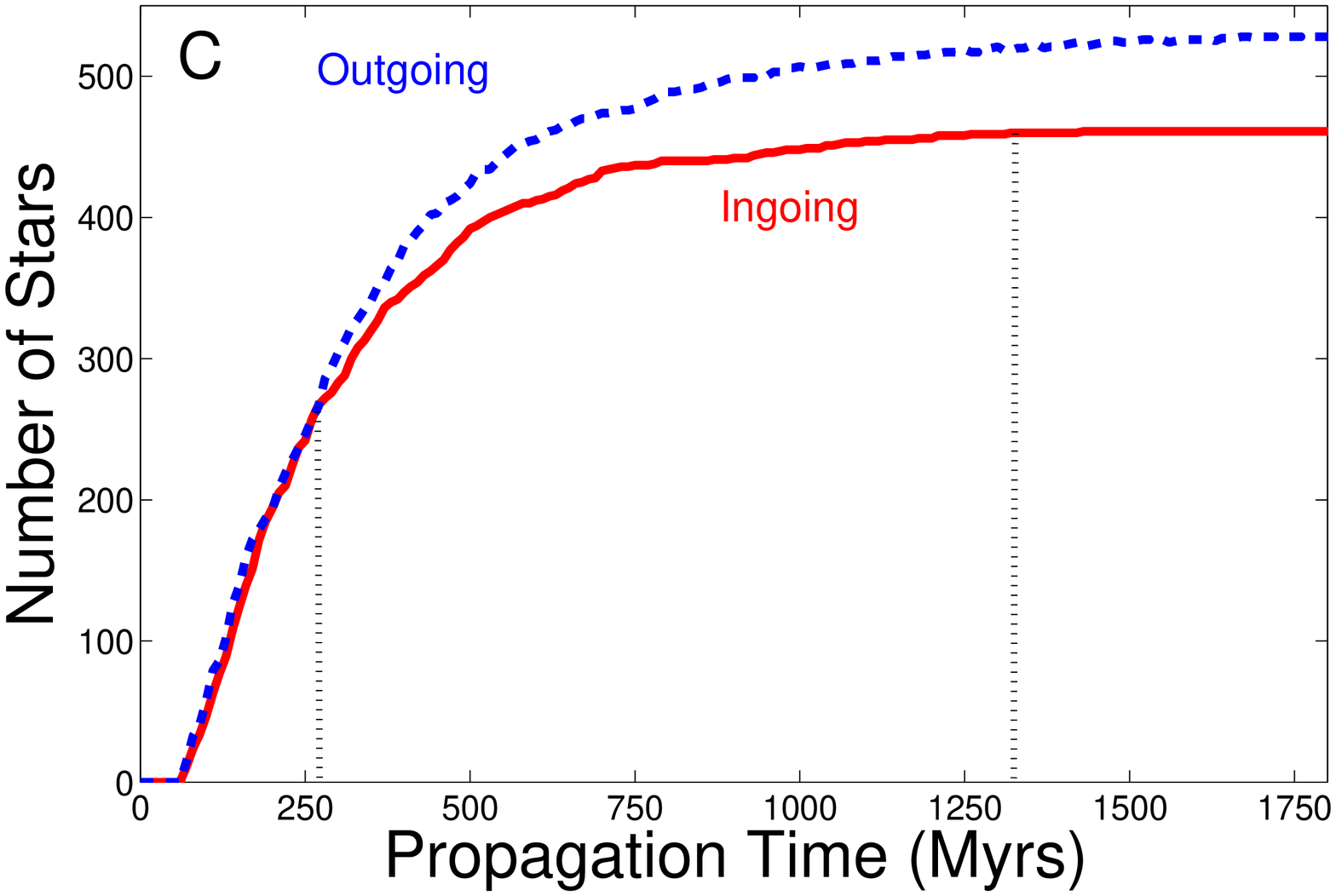}\tabularnewline
\end{tabular}

\caption{\label{f:MS-HVSs-asym} The asymmetric distribution of outgoing vs.
ingoing stars. The curves show the integrated number of outgoing vs.
ingoing stars below (or above, respectively) the critical asymmetry
lines (PAC model), corresponding to the range of propagation times
in the Galactic potential. The asymmetry begins at propagation times
of $\sim2.5\times10^{8}$ yrs (see left dotted vertical line). No
ingoing stars are observed to correspond to propagation times beyond
$\sim1.3\times10^{9}$ yrs.}

\end{figure}

In this calculation we assumed a specific maximal propagation time
for the GC HVSs. Instead, we can look at the asymmetric distributions
for stars that propagated in the Galactic potential for shorter propagation
times. Such stars could be either more massive MS B stars with shorter
MS lifetimes, or MS stars ejected from the GC only after evolving
for some time in the GC (or ejected more recently from the GC). Since
the frequency of more massive stars is small, the latter possibilities
are more likely to apply. One would therefore expect to see an asymmetric
distribution even at times shorter then the MS lifetime of $3\, M_{\odot}$
B MS star ($4\times10^{8}$ yrs), likely of the order of half this
time ($\sim2\times10^{8}$ yrs), assuming a continuous ejection rate
of HVSs. In fig. \ref{f:MS-HVSs-asym} we show the number of outgoing
vs. ingoing stars for a range of propagation times. For the PAC potential
model we used, the distribution begins to become asymmetric at propagation
times of $2.5\times10^{8}$ yrs, in good agreement with our expectations. 

For completeness we discuss the possibility that the HVSs are not
MS stars but hot BHB stars ejected from the GC. In this case their
absolute magnitude is different from that of MS stars, and the inferred
distances change accordingly (see fig. \ref{f:MS-HVSs}b). Again we
can look for the typical propagation time at which we see an asymmetric
distribution of these stars assuming they were ejected from the GC.
We find this to be at a few $10^{8}$ yrs. This time is comparable,
but longer, than the lifetime of hot BHB stars at this phase. It is
also much shorter than the lifetime of hot BHB progenitors, which
could extend up to a few Gyrs. Both of these inconsistencies, together
with the dynamical constraints against the ejection of hot BHB HVSs
\citep{per08a} suggest a different identification of these stars.
Furthermore, comparing their velocity-distance distribution to that
of regular BHB stars in the halo \citep{xue+08} show very similar
distribution (see fig. \ref{f:MS-HVSs}b), i.e. the observed cut-off
in the distribution is not related to the critical lines we described,
but due to the limit of the symmetric distribution of halo objects.

Another possibility is that asymmetries in the HVSs candidates sample
are not due to the population of HVSs, but due to stellar streams
located at specific regions of the Galactic halo, which could have
correlated velocities, and not an isotropic distribution as we assumed.
To check this we repeated the asymmetry calculations described above,
but this time for several different and distinct regions in the Galactic
halo (different Galactic longitudes). Although some differences in
the strength of the asymmetry are observed, all regions showed a clear
asymmetry bias towards outgoing stars.

\section{Summary}

\label{sec:summary}

In this paper we studied the characteristics, the origins and the
use of stars observed in the survey for HVSs in the Galactic halo.
The kinematics of currently observed HVSs ejected from the Galactic
center depend strongly on their lifetimes and their propagation in
the Galactic potential. We suggest a novel method to probe the Galactic
potential up to large distances using the kinematics and the spectral
identification of HVSs. We also use a reverse method, where a specific
Galactic potential model is assumed, to give lower limit estimates
on the number of HVSs ejected from the Galactic center. Future observations
of HVSs in M31 \citep{she+07} and in other Galaxies (Perets et al.,
in preparation) could have a similar use for studies of Galactic structures
and potentials.

\appendix{}

\section{Galactic potential models}

In the following we shortly discuss the Galactic potential models
used in section \ref{sub:potentials} (KZS, BSC, RAVE1-2, PAC and
MOND models) and the differences between them. For CDM models, we
test \citet{kly+02} B1 model (KZS model), with a double-exponential
disc for baryons and a NFW profile \citep{nav+96} for a dark matter
halo. We also replace baryons with the Besancon model together with
the same CDM component in KZS B1 (BSC model). After that we adopt
two of the models appeared in RAVE survey \citep{smi+07}. In both
RAVE models we apply \citet{miy+75} disc and Hernquist bulge \citep{her90}
for baryons, while for CDM, one of the model is uncontracted NFW (RAVE
1 model) and the other has a \citet{wil+99} profile (RAVE 3). The
PAC model uses \citet{miy+75} potential for the disk, bulge and halo
components parametrized to match observations (see \citealt{pac+90}
for details).

The KZS and BSC models we use have the same NFW dark matter profile,
but they differ in the distribution of their baryonic component. However,
the propagation of the HVSs is mostly dominated by the cold dark matter
halo far in the halo, and is therefore quite similar in these different
models, as well as to the PAC model. The RAVE 1 model also uses the
NFW profile, but with a different set of parameters (see \citealt{smi+07}
for details). In this model the dark matter halo is much more concentrated
than the KZS/BSC \citep{kly+02} models and the virial mass is almost
twice as large as the virial mass in the KZS/BSC. The RAVE-3 model
has a steeper distribution towards the center ($\rho\propto r^{-2}$),
and a sharp cut-off at large radii. The MOND model uses the same baryon
density of BSC but the potential in MOND is deeper than any of the
CDM models (with the strength of external field $0.01\, a_{0}$).
When the external field is stronger, up to $0.03\, a_{0}$ (corresponding
to the upper limit from the constraints of the RAVE survey), the MONDian
potential becomes very similar to the KZS/BSC models up to the distances
of the observed HVSs, and differ significantly only at larger radii.
Due to the similarity between some of the the models at the distances
of the observed HVSs, it is difficult to discriminate between the
potentials of the KZS/BSC, PAC and MOND using the critical asymmetry
lines we present.

\begin{figure}
\begin{tabular}{c}
\includegraphics[clip,width=0.45\columnwidth]{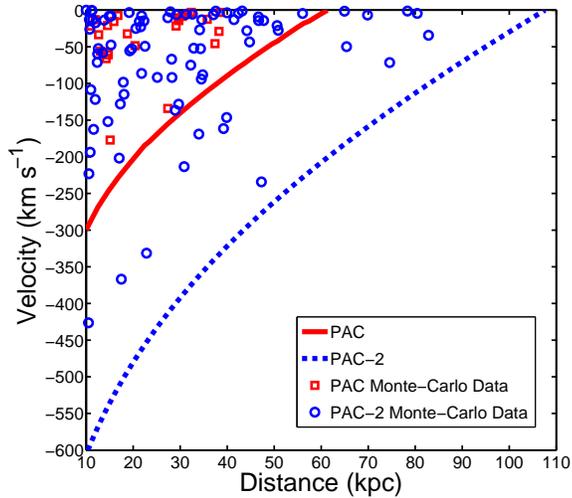}\tabularnewline
\end{tabular}

\caption{f\label{f:mock-hvss} The distribution of ingoing simulated HVSs.
The squares represent HVSs propagating in the PAC model, and the circles
represent HVSs propagating in the PAC-2 model (see text). The critical
asymmetry lines are shown for both Galactic potential models, PAC
(solid line) and PAC-2 (dashed line).}

\end{figure}

\section{Constraining the Galactic potential: example of simulated data}

In the following we use a simple example of mock simulated data of
HVSs to show the way in which galactic potentials could be constrained
by future HVSs data. Using Monte-Carlo simulations of HVSs propagating
in various galactic potential models we produce distance-velocity
plots of their expected distributions. This mock data is then compared
with the critical asymmetry lines corresponding to the different models. 

In this example we assume the HVSs were ejected from the Galactic
center following binary disruption by the MBH. We make use of the
same methods described by \citet{bro+06c} and \citet{ken+08} to
simulate the HVSs ejection velocities, assuming the progenitor binaries
of the HVSs were $3+3\, M_{\odot}$ binaries. For each ejected star
we choose a random ejection time during its lifespan, and a random
ejection time during the last $4\times10^{8}$ yrs (the lifespan of
$3\, M_{\odot}$ stars; no such star could survive to current day
if it were ejected earlier). We reject those stars which would end
their lifespan on the main sequence before current day, i.e. they
would not have been observable today. Each of the potentially observable
HVSs is then propagated in a given Galactic potential until the current
day. For simplicity we show the results for the propagation of HVSs
in two Galactic potential models, which differ in only one parameter.
We use the PAC model which is parametrized by eight paramters (see
details in \citealp{pac+90}), to produce two potentials. The first
(PAC) is the Galactic potential as originally parametrized in \citet{pac+90},
and the second (PAC-2) differs only in the halo core radius, $r_{c},$which
is taken to be $r_{c}=2\, kpc$ instead of $r_{c}=6\, kpc$ in the
original PAC model. The number of simulated stars was chosen such
that the total number of HVSs with velocity greater than $450\, km\, s^{-1}$
observable in the Galactic halo would be $\sim100$, i.e the total
number of such HVSs estimated to exist in the Galaxy based on current
surveys. 

The distance-velocity distribution of ingoing simulated HVSs is shown
in fig. \ref{f:mock-hvss}, together with the critical asymmetry lines
for the two models. As can be seen in the figure, HVSs propagating
in the PAC-2 model obtain distance-velocity position beyond the critical
asymmetry line of the PAC model, i.e. the PAC model would be directly
excluded, independently of the detailed distribution of the HVSs.
HVSs propagating in the PAC model could never get beyond the critical
asymmetry line of the PAC-2 model and therefore can never totally
exclude this model. Nevertheless, the probability for not observing
even a single HVS beyond the PAC critical asymmetry line, assuming
that the HVSs did propagate in the PAC-2 model is $\sim0.007$ (one
would expect to see 5 such stars in the sample of 25 ingoing HVSs,
given the PAC-2 model data, but none is observed), i.e. rejecting
the PAC-2 model at high confidence. 

\acknowledgements{We would like to thank Warren Brown for helpful discussions and for
supplying us with the data on the kinematics of the stars in the HVSs
sample. HBP would also like to thank the Israeli Commercial \& Industrial
Club for their support through the Ramon scholarship. BF is a postdoctoral
researcher of the FNRS (Belgium). GG is a postdoctoral fellow with
the National Science Fund (FWO-Vlaanderen). TA is supported by ISF
grant 928/06 and ERC Starting Grant 202996 and a new Faculty grant
by sir H. Djangoly, CBE, of London UK. }

\bibliographystyle{apj}

\end{document}